\documentclass[11pt]{article}
\pdfoutput=1
\usepackage{jheppub}

\usepackage[svgnames,table]{xcolor}
\usepackage{graphicx}
\usepackage{amsfonts,amsmath,amssymb,amsthm}
\usepackage{mathrsfs,bm,bbm,esint}
\usepackage[mathscr]{eucal}
\usepackage{physics}
\usepackage{slashed,soul}
\usepackage{rotating,pdflscape}
\usepackage{enumerate}
\usepackage{multirow,array}
\usepackage[numbers,sort&compress]{natbib}
\usepackage{tikz}
\usetikzlibrary{cd}
\usetikzlibrary{decorations.pathmorphing}
\usetikzlibrary{decorations.pathreplacing}
\usetikzlibrary{decorations.markings}
\tikzset{snake it/.style={decorate, decoration=snake}}
\tikzset{->-/.style={decoration={
  markings,
  mark=at position .5 with {\arrow{>}}},postaction={decorate}}}
\usepackage{subcaption}
\usepackage{float}
\usepackage{afterpage}
\usepackage{cleveref}
\crefname{section}{\S\!\!}{\S\S\!\!}
\Crefname{section}{\S}{\S\S}
\crefname{appendix}{Appendix}{Appendices\!}
\crefname{figure}{Fig.\!}{Figs.\!}

\definecolor{rust}{rgb}{0.8,0.2,0.2}

\newcommand{\half}{\frac{1}{2}}

\newcommand{\Mbb}{\mathscr{M}}
\newcommand{\rhob}{\hat{\rho}}
\newcommand{\sigmab}{\hat{\sigma}}
\newcommand{\Ob}{\hat{\mathcal{O}}}
\newcommand{\Idop}{{\bm 1}}

\title{Petz reconstruction in random tensor networks}

\author{Hewei Frederic Jia}
\author{, Mukund Rangamani}
\affiliation{Center for Quantum Mathematics and Physics (QMAP)\\
Department of Physics, University of California, Davis, CA 95616 USA}

\emailAdd{fjia@ucdavis.edu}
\emailAdd{mukund@physics.ucdavis.edu}

\abstract{
We illustrate the ideas of bulk reconstruction in the context of random tensor network toy models of holography. 
Specifically, we demonstrate how the Petz reconstruction map works to obtain bulk operators from the boundary data by exploiting the replica trick.  We also take the opportunity to comment on the differences between coarse-graining and random projections.
}

\begin{document}
\maketitle


\section{Introduction}
\label{sec:intro}
 
The emergence of bulk spacetime geometry from non-gravitational field theoretic degrees of freedom in the  AdS/CFT correspondence can be understood by viewing the holographic map from the bulk to the boundary as a quantum error correcting code \cite{Almheiri:2014lwa,Dong:2016eik}. The essential idea is that while the Hilbert space of the field theory is isomorphic to the full string theoretic quantum gravitational Hilbert space, semiclassical gravitational physics has access to a much smaller subspace of states. These `code subspace' states corresponding to  excitations of the vacuum (or other geometric states) by a few, $\order{c_\text{eff}}$, perturbative quanta, are to be viewed as the quantum message one wishes to encode into a bigger Hilbert space.\footnote{ We use $c_\text{eff}$  to denote the effective central charge of the field theory. In the gravitational setting it can be related to the AdS scale in Planck units: $c_\text{eff}  \sim \left(\frac{\ell_\text{AdS}}{\ell_P}\right)^{d-1}$. In the familiar AdS$_5\! \times {\bf S}^5$ example it is related to the rank of the gauge group, $c_\text{eff} \sim N^2$.} This encoding map can moreover be viewed as a noisy quantum channel. 

The question of recovering local bulk geometry can be rephrased in this framework as constructing a recovery map for this channel, one that allows us to reconstruct from field theory data, either the bulk state, or better yet (in the Heisenberg picture) local bulk operators. The latter are especially interesting given that the standard reconstruction of local bulk physics exploits only the bulk causal structure \cite{Banks:1998dd,Hamilton:2006az} through the extrapolate dictionary \cite{Harlow:2011ke,Heemskerk:2012mn}. It however has become quite clear thanks to the holographic entanglement entropy proposals \cite{Ryu:2006ef,Hubeny:2007xt} that one should be able to reconstruct operators in a larger domain of the bulk, the entanglement wedge  \cite{Czech:2012bh,Headrick:2014cta,Wall:2012uf}.

While it has been argued that reconstructing operators in the entanglement wedge involves modular evolution \cite{Jafferis:2015del,Faulkner:2017vdd}, an alternative viewpoint exploiting quantum recovery maps was presented in \cite{Cotler:2017erl}. The idea as elaborated further in   \cite{Chen:2019gbt} is that the Petz map \cite{Petz:1986tvy} and its twirled generalization \cite{Junge:2015lmb}, provide a universal general-purpose recovery maps which suffice to reproduce the bulk quantum state with fidelity $\mathcal{F}\sim 1-\order{c_\text{eff}^{-\frac{1}{2}}}$.  Explicit reconstructions have been analyzed in \cite{Almheiri:2017fbd,Chen:2019iro} using modular flow, and  in \cite{Penington:2019kki} using the  Petz map. Some of these discussions have explicitly demonstrated the non-trivial encoding of the bulk in the boundary, especially in the context of black hole evaporation through the replica wormhole contributions \cite{Penington:2019kki,Almheiri:2019qdq} (which justify the quantum extremal surface \cite{Engelhardt:2014gca}  prescription). While there are structural similarities between the modular evolved operators, and the Petz reconstruction, the precise connection between the two is as yet to be fully fleshed out.

The goal of this short note is to explore the properties of the Petz map in a simple toy model of holography, viz., random tensor networks (RTN) \cite{Hayden:2016cfa}. These tensor networks involve discrete degrees of freedom with the bond dimension $\chi$ a proxy for the central charge, $c_\text{eff}\sim \log\chi$. Unlike specific perfect tensor codes which work with fixed set of operators/tensors \cite{Pastawski:2015qua}, the random tensors involve projecting onto suitably entangled states (analogous to PEPS tensor networks, cf., \cite{Orus:2013kga}). As demonstrated in \cite{Hayden:2016cfa} and argued for more generally in \cite{Harlow:2016vwg} these discrete networks share many features of the holographic map, in particular, saturating the minimal cut rule for computing von Neumann entropy, analogous to the RT/HRT formulae in gravitational systems \cite{Ryu:2006ef,Hubeny:2007xt}. One can moreover understand the flat entanglement spectra of these models by invoking a constrained variational problem in gravity, wherein one works with fixed-area states of the bulk gravitational path integral \cite{Akers:2018fow,Dong:2018seb}. Our motivation is to provide a simple example (analogous to the discussion in \cite{Penington:2019kki}) for the Petz reconstruction. In addition we will comment on some of the features of the RTNs and holographic codes viewed as quantum channels. 

The outline of this article is as follows: in \cref{sec:rtnrep} we give a quick overview of the RTNs and the use of the replica trick to compute entropies. In \cref{sec:petzb} we illustrate how the Petz reconstruction works in the case of the RTNs, demonstrating along the way, the use of replicas to show the matching of bulk and boundary observables. Finally, in \cref{sec:comments} we comment on various features of holographic encodings of the bulk geometry viewed from the perspective of quantum channels.

\section{Random tensor networks and replicas}
\label{sec:rtnrep}

%
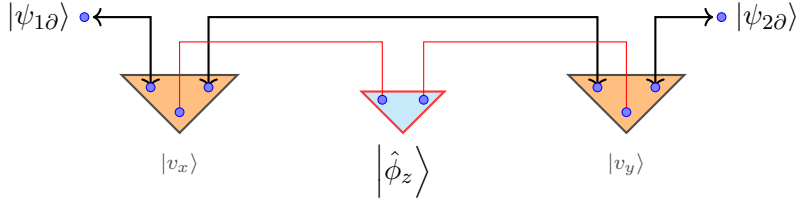
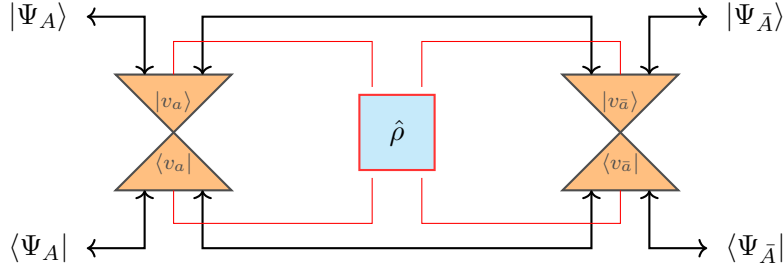
\begin{figure}[h]
\begin{subfigure}{\textwidth}
\centering
\begin{tikzpicture}
[scale=1.1,
tzop/.style={rectangle,draw=red!80,fill=cyan!20,thick, inner sep=0pt,minimum size=10mm},
tzstates/.style={black!70,fill=orange!50,thick, inner sep=0pt,minimum size=5mm},
tproj/.style={black,thick,<->},
tbulk/.style={red,thin},
]
\draw[tzop]  (-0.5,0.5)  -- ++(1,0) -- ++(-0.5,-0.5)  node[below] {$\ket{\hat{\phi}_z}$}  -- cycle;
\draw[tzstates] (-2.7,0) node[below=1.5mm] (k1) {$\scriptstyle{\ket{v_x}}$} -- ++(0.7,0.7) -- ++(-1.4,0) -- cycle;
\draw[tzstates] (2.7,0)   node[below=1.5mm] (k2) {$\scriptstyle{\ket{v_y}}$} -- ++(0.7,0.7) -- ++(-1.4,0)  -- cycle;
\draw[tproj] (-2.35,0.55) -- ++ (0,0.85) -- ++(4.7,0) -- ++(0,-0.85);
\draw[tproj] (-3.05,0.55) -- ++ (0,0.85) -- ++(-0.7,0) node[left=1.5mm] {$\ket{\psi_{1\partial}}$};
\draw[tproj] (3.05,0.55) -- ++ (0,0.85) -- ++(0.7,0) node[right=1.5mm] {$\ket{\psi_{2\partial}}$};
\draw[tbulk]  (-2.7,0.25) -- ++ (0,0.85) -- ++(2.45,0) -- ++(0,-0.7);
\draw[tbulk]  (2.7,0.25) -- ++ (0,0.85) -- ++(-2.45,0) -- ++(0,-0.7);
\draw[color =blue, fill=blue!50] 
	(-3.05,0.55) circle [radius=1.5pt] 
	(-2.35,0.55) circle [radius=1.5pt]
	(-2.7,0.25) circle [radius=1.5pt] 
	(-0.25,0.4) circle [radius=1.5pt] 
	(0.25,0.4) circle [radius=1.5pt] 
	(2.35,0.55) circle [radius=1.5pt] 
	(2.7,0.25) circle [radius=1.5pt] 
	(3.05,0.55) circle [radius=1.5pt] 
	(3.85,1.4) circle [radius=1.5pt] 
	(-3.85,1.4) circle [radius=1.5pt] 
	;
\end{tikzpicture}
\caption{The map from the bulk to boundary in a random tensor network illustrated for three vertex graph. At sites $x$ and $y$ we have three qubits (blue circles) while at site $z$ we have two qubits. In addition we have two boundary qubits shown at the dangling ends. The black lines with arrowheads between qubits denote an EPR state in the tensor product space of the linked qubits. At vertex $z$ we place a state $\ket{\hat{\phi}_z}$ of the two qubits while at vertices $x$ and $y$ we include  random states $\ket{v_x}$ and $\ket{v_y}$, respectively of the three qubits located there. These are collectively encoded by the  triangles: a blue one stands  for operator insertion and an orange one encodes random state of the qubits. The red lines indicate contractions between the qubits where we insert operators to those at the other vertices. The network defines a map from the bulk qubits to the boundary qubits piping the state $\ket{\hat{\phi}_z}$ onto a state in $\ket{\psi_{1\partial}}\otimes \ket{\psi_{2\partial}}$.}
\label{fig:rtnmapqubits}
\end{subfigure}

\begin{subfigure}{\textwidth}
\centering
\begin{tikzpicture}
[scale=1.1,
tzop/.style={rectangle,draw=red!80,fill=cyan!20,thick, inner sep=0pt,minimum size=10mm},
tzstates/.style={black!70,fill=orange!50,thick, inner sep=0pt,minimum size=5mm},
tproj/.style={black,thick,<->},
tbulk/.style={red,thin}
]
\node at (0,0) [tzop]  {$\rhob$} ;
\draw[tzstates] (-2.7,0) node[below=1.5mm] (b1)  {$\scriptstyle{\bra{v_a}}$} node[above=1.5mm] (k1) {$\scriptstyle{\ket{v_a}}$} -- ++(0.7,0.7) -- ++(-1.4,0) -- ++(1.4,-1.4) -- ++(-1.4,0) -- cycle;
\draw[tzstates] (2.7,0) node[below=1.5mm] (b2)  {$\scriptstyle{\bra{v_{\bar{a}}}}$} node[above=1.5mm] (k2) {$\scriptstyle{\ket{v_{\bar{a}}}}$} -- ++(0.7,0.7) -- ++(-1.4,0) -- ++(1.4,-1.4) -- ++(-1.4,0) -- cycle;\draw[tproj] (-2.35,0.7) -- ++ (0,0.7) -- ++(4.7,0) -- ++(0,-0.7);
\draw[tproj] (-2.35,-0.7) -- ++ (0,-0.7) -- ++(4.7,0) -- ++(0,0.7);
\draw[tproj] (-3.05,-0.7) -- ++ (0,-0.7) -- ++(-0.7,0) node[left=1mm] {$\bra{\Psi_A}$};
\draw[tproj] (-3.05,0.7) -- ++ (0,0.7) -- ++(-0.7,0) node[left=1mm] {$\ket{\Psi_A}$};
\draw[tproj] (3.05,-0.7) -- ++ (0,-0.7) -- ++(0.7,0) node[right=1mm] {$\bra{\Psi_{\bar{A}}}$};
\draw[tproj] (3.05,0.7) -- ++ (0,0.7) -- ++(0.7,0) node[right=1mm] {$\ket{\Psi_{\bar{A}}}$};
\draw[tbulk]  (-2.7,0.7) -- ++ (0,0.4) -- ++(2.4,0) -- ++(0,-0.55);
\draw[tbulk]  (-2.7,-0.7) -- ++ (0,-0.4) -- ++(2.4,0) -- ++(0,0.55);
\draw[tbulk]  (2.7,0.7) -- ++ (0,0.4) -- ++(-2.4,0) -- ++(0,-0.55);
\draw[tbulk]  (2.7,-0.7) -- ++ (0,-0.4) -- ++(-2.4,0) -- ++(0,0.55);
\end{tikzpicture}
\caption{ We can abstract the information of the network into a few essential elements to indicate the bulk to boundary map. We first extend \cref{fig:rtnmapqubits} to include the conjugate bra states so as to denote maps from bulk operators/states $\rhob$ onto boundary operators $\rho_{A\bar{A}}$. In the process we concatenate operator insertion vertices into a square (blue), while the vertices with random projectors $\Pi_V$ are depicted  by orange triangles. The black lines  continue to denote the maximally entangled state between subfactors, red lines denote contractions between the subfactors where we insert the operator and those where we perform random projections. Dangling lines at the outer edge lead  to boundary degrees of freedom. In this presentation, for a boundary bipartitioning we can truncate to three effective bulk vertices, one where the operator $\rhob$ is inserted and two others  ($a$ and $\bar{a}$) which correspond to the two boundary degrees of freedom ($A$ and $\bar{A}$).}
\label{fig:rtnmapabstract}
\end{subfigure}
\caption{An illustration of the random tensor network.}
\label{fig:rtnmap}
\end{figure}

We begin with a quick overview of RTNs \cite{Hayden:2016cfa}.  Consider an arbitrary graph with a vertex set $\{x\}$. At each vertex we have a Hilbert space $\mathcal{H}_x =\otimes_{k=1}^{n_x} \mathcal{H}_{x,k}$ which admits a tensor product decomposition into factors $\mathcal{H}_{x,k}$. For two vertices $x$ and $y$ that are connected by a link, we pick some sub-factors  of the vertex Hilbert spaces, and use them to define a link Hilbert space $\mathcal{H}_{xy} = \left(\otimes_{k=1}^{m_x} \mathcal{H}_{x,k}\right) \otimes \left(\otimes_{l=1}^{m_y} \mathcal{H}_{y,l}  \right)$ with $m_x \leq n_x$ and $m_y \leq n_y$, respectively. In addition we will allow for  some boundary vertices which lie at the edge of the graph and have a Hilbert space $\mathcal{H}_\partial$. These will likewise have links to some bulk vertices with a similar construction defining $\mathcal{H}_{x\partial}$.

Given this set-up we can define the tensor network in two equivalent ways. We first lay out maximally entangled states $\tau_{xy}$ in the link Hilbert space $\mathcal{H}_{xy} \otimes \mathcal{H}_{yx}$  along each link including the dangling boundary ones. We  additionally choose to insert a bulk operator $\Ob$ (or state $\rhob$) on a set of bulk vertices by acting with an appropriate operator on a collection of vertices.\footnote{ Bulk operators are denoted with a hat for clarity.} Finally,  on the state thus prepared on $\otimes_{x} \mathcal{H}_x$, we make random  measurements at each bulk vertex by projecting onto Haar random state $\ketbra{V_x}{V_x}$. We can obtain the state $\ket{V_x}$ by acting on a reference state in $\mathcal{H}_x$ with a Haar random unitary, eg., $\ket{V_x} = U_x \ket{0}$. A basic three node network is illustrated in \cref{fig:rtnmapqubits}.

 We will refer to the  tensor indices associated with the vertex set in the bulk of the graph where we insert operators as ``bulk degrees of freedom'' while those corresponding to uncontracted, dangling indices from the vertex set are our  ``boundary degrees of freedom''.   Operationally we can take the tensor factors for the bulk Hilbert space to be smaller in dimension compared to those where we insert the random projectors (to allow for saddle point analysis). The network then defines a map from the bulk to the boundary, mapping bulk quantum states $\rhob$ onto boundary states via
\begin{equation}\label{eq:bulkbdymap}
\begin{split}
\Mbb(\rhob) &= \Tr_\text{bulk}\left(\rhob \; \Pi_V \ketbra{\Phi}{\Phi} \right)\,, \\
\Pi_V = \otimes_x \, \ketbra{V_x}{V_x} \,,\quad  & \qquad\quad \ket{\Phi} = \otimes_{\langle xy\rangle}\ket{\tau_{xy}}\,,
\end{split}
\end{equation}	
where $\Pi_V$ is our random measurement.

An alternative perspective is to first start with random states prepared at each vertex in $\mathcal{H}_x$, and thence project them onto the maximally entangled state along the links. The construction of the network and the boundary states is illustrated in Fig.~\ref{fig:rtnmap}. We will view the network as preparing a quantum state with no information about temporal evolution. Consequently, our statements should be viewed as being applicable to states on a single Cauchy slice in the geometric set-up.

We will bipartition the boundary degrees of freedom into $A$ and $\bar{A}$. Associated with these will be a set of bulk vertices, collectively $a$ and $\bar{a}$, respectively, with $\dim(\mathcal{H}_a) = d_a$ and $\dim(\mathcal{H}_{\bar{a}}) = d_{\bar{a}}$. The bulk Hilbert space is viewed as  the code subspace of the boundary and has dimension $d_\text{code} = d_a \times d_{\bar{a}}$.  In the geometric setting the state space $\mathcal{H}_a$ would correspond to bulk states in the homology surface $R_A \equiv a$, a Cauchy slice of the entanglement wedge $\mathcal{E}_A$ of $A$. The map $\Mbb(\rhob)$ and its restriction to a subregion of the boundary $\Mbb_A({\rhob}) \equiv \Mbb(\rhob)|_A$ (obtained by partial tracing) are not normalized, i.e., the maps are not trace-preserving. We will choose to convert the outputs to boundary density operators, normalizing by hand, to obtain $\rho$ and $\rho_A$, respectively. This will lead to a normalization factor of $\Tr(\Mbb(\rhob))$ in the computations below. We will revisit the nature of the bulk to boundary map later in our discussion.

Let us quickly review the computation of  von Neumann and R\'enyi entropies for $\rho_A$  using the replica method, cf., \cite{Hayden:2016cfa} for details. We compute traces of powers of $\rho_A$ (with aforementioned normalization) by working in the covering space unfolding the powers of reduced density operators to the computation of the expectation value of an observable, viz., 
\begin{equation}\label{eq:rtnreplica1}
\Tr_A(\rho_A^n) = \frac{\Tr_A[\Mbb_A^n(\rhob)]}{\Tr[\Mbb(\rhob)]^n} =  \frac{\Tr[\Mbb(\rhob)^{\otimes n} \, X^{(n)}_A ]}{\Tr[\Mbb(\rhob)^{\otimes n}]}\,.
\end{equation}
The first equality is the definition, and in writing the second we used the fact that the trace over a set of powers of an operator can be computed by working in an $n$-fold tensor product, together with a suitable projection on the symmetric subspace achieved by the insertion of $X_A^{(n)}$. On the $n$-fold tensor product we have a natural $S_n$ permutation group action, which for the purposes of computing cyclic  trace invariants introduces a cyclic permutation of $n$ elements.  This defines the operator $X^{(n)}_A =\prod_{x\in A} z_{(n)}$ above, with $z_{(n)} \in \mathbb{Z}_n$ a cyclic permutation. Note that the cyclic permutation is inserted in the boundary region $A$ alone, since the $\bar{A}$ has already been traced over to compute $\rho_A$.

The main advantage of the RTNs is that the replica calculation can be done quite efficiently by first averaging over the random projectors.\footnote{ We use overbars to denote both the complementary regions in the bipartition ($A$ and $\bar{A}$) as well as the average over random projectors, which hopefully will not cause confusion; the distinction should be apparent from the context.}  
We can use the result:\footnote{${\sf N}_{n,x} = \frac{(d_x + n-1)!}{(d_x-1)!}$ is a normalization factor depending on $\dim(\mathcal{H}_x)  \equiv d_x$, but since it will cancel out with our normalization of $\rho_A$, we will not write it out explicitly.}  
\begin{equation}\label{eq:projavg}
\overline{\ketbra{V_x}{V_x}^{\otimes n}}  \equiv \overline{\left(U_x \ketbra{0}{0} U_x^\dagger\right)^{\otimes n}}  = \int [DU] \left(U_x \ketbra{0}{0} U_x^\dagger\right)^{\otimes n} =
 {\sf N}_{n,x}\, \sum_{g_x \in S_n} \, g_x \,.
\end{equation}	
This ends up mapping the problem to a spin model with $S_n$ valued spins $g_x$ at each vertex.  The numerator of \eqref{eq:rtnreplica1} maps to the computation of partition function, with boundary conditions: insertions of fixed cyclic permutation $X_A^{(n)}$ along boundary spins in $A$ and the identity element $e$ in $\bar{A}$. Therefore, 
\begin{equation}\label{eq:rtnreplica2}
\begin{split}
\overline{\Tr[\Mbb(\rhob)^{\otimes n} \, X_A^{(n)}] }
&= 
	{\sf N}_n \, \sum_{g_x \in S_n} \!  {}^{\otimes n}\!\!\mel{\Phi}{g\, z_{(n),A}}{\Phi}^{\otimes n} \, \Tr\left(\rhob^{\otimes n}\,g\right) \\
&=
	{\sf N}_n\, e^{-(n-1) \, \abs{\gamma_A}\,\log\chi  }	\Tr(\rhob_a^n) + \cdots .
\end{split}
\end{equation}
where $g = \prod_x g_x$, ${\sf N}_n=\prod_x {\sf N}_{n,x} $,  $\chi$ is the bond dimension, and ellipses stand for subleading  terms.  $\gamma_A$ the minimal cut of the graph implementing a bulk bipartitioning into $a$ and $\bar{a}$.

Since we are computing  a spin-model partition sum with fixed boundary conditions, we can estimate the result for large bond dimensions, by thinking of spin domains. For our boundary conditions $g_x = z_{(n)}  \in \mathbb{Z}_n$ for $x \in A$ and $g_x =e$ for $x \in \bar{A}$, we propagate the boundary spins inwards into the graph and encounter a domain wall separating the two domains in an energy minimizing configuration. Generically, there will be multiple competing configurations with domain walls $\gamma_A^i$, serving as the bulk separatrix (the RTN analog of the bulk RT surfaces). Energy minimization picks out a unique preferred configuration in the spin model \cite{Hayden:2016cfa}, and thus in the RTN serves to define the homology surface $a$ and $\bar{a}$, for $A$ and $\bar{A}$, respectively. As one changes the relative dimensions of $\mathcal{H}_A$ and $\mathcal{H}_{\bar{A}}$ we will encounter phase transitions with the minimum energy domain wall configuration switching between competing saddles. The ellipses in \eqref{eq:rtnreplica2} refer to the contribution of these subleading saddles and are of order $\mathcal{O}(\chi^{\abs{\gamma_A^1} - \abs{\gamma_A^2}})$.

  The normalization factor in the denominator evaluates similarly though now all boundary spins have $g=e$ permutation, resulting in $\Tr[\Mbb(\rhob)^{\otimes n}] = {\sf N}_n + \cdots$,
leading to
\begin{equation}\label{eq:rtnrenyi}
S_A^{(n)} = \frac{1}{1-n}\, \log\Tr\left(\rho_A^n\right) =\abs{\gamma_A} \,  \log \chi + S^{(n)}_a\,.
\end{equation}	
The entanglement spectrum of these networks is flat (i.e., $n$ independent) \cite{Hayden:2016cfa} and can be understood to correspond  to fixed-area states in the geometric description \cite{Akers:2018fow,Dong:2018seb}.

\section{Petz reconstruction of bulk states}
\label{sec:petzb}

The goal of bulk reconstruction is to construct an operator $\mathcal{O}_A$ supported on $A$, given a bulk operator $\Ob_a $ supported on the homology surface $R_A =a$ of $A$. We can view the state $\rho_A$ obtained from $\rhob$ as the result of operating a noisy quantum channel, $\rho_A = \mathscr{E}(\rhob)$. Our task is to construct a recovery map $\mathscr{R}$, such that $\mathscr{R}\circ \mathscr{E}(\rhob) = \mathscr{R}(\rho_A)\approx \rhob_a$.  We then can use the adjoint channel $\mathscr{R}^\dagger$ to find a map between boundary and bulk operators, for $\mathcal{O}_A = \mathscr{R}^\dagger(\Ob_a)$. 

For general quantum channels the Petz map gives an ansatz for this recovery: in the Heisenberg picture for operators it reads
\begin{equation}\label{eq:petz}
\mathscr{O}_A  = \mathscr{E}(\sigma)^{-\half} \, \mathscr{E}\left(\sigma^{\half} \, \Ob \, \sigma^{\half} \right) \mathscr{E}(\sigma)^{-\half}
\end{equation}	
where $\sigma$ is a fixed fiducial state. Taking it to be the maximally mixed state $\tau$ achieves the desired reconstruction with error of $\order{c_\text{eff}^{-\half}}$ \cite{Cotler:2017erl,Chen:2019gbt}. For starters however, we will employ a simpler ansatz, dubbed ``Petz-lite'' in \cite{Penington:2019kki} which posits instead
\begin{equation} 
\mathcal{O}_A \propto \mathcal{E}(\Ob) \,,
\end{equation}
with the coefficient fixed by demanding that the identity operator maps to the identity operator.

One potential obstruction in using the Petz map in RTNs is the non-linearity in the bulk-boundary map \eqref{eq:bulkbdymap} arising from the fact that we have to normalize the boundary state $\rho_A$ by hand. Strictly speaking, we are dealing with a general quantum operation since the map from bulk to boundary is not trace-preserving. We nevertheless can use the map $\Mbb_A$ to construct analogs of Petz like reconstruction maps for operators. We will first explain how to perform the reconstruction using the Petz-lite ansatz and then talk about the more general twirled Petz map.

\subsection{The simplified Petz reconstruction}
\label{sec:petzlite}

For the simple reconstruction, we claim that the map from the bulk to the boundary $\Mbb_A$ will itself suffice to perform the reconstruction, viz., 
\begin{equation}\label{eq:petzlite}
\mathcal{O}_A = c_0\, \Mbb_A(\Ob) \,,
\end{equation}	
provides a faithful boundary representative of a bulk operator in $a$. Unlike density operators we will not a-priori normalize the pull-back of the bulk operators to the boundary, but will determine the proportionality constant $c_0$ post-facto so that 1-point functions agree.

\begin{figure*}[ht]
\centering
\begin{tikzpicture}
[scale=1.0,
tzop/.style={rectangle,draw=red!80,fill=cyan!20,thick, inner sep=0pt,minimum size=10mm},
tzstates/.style={black!70,fill=orange!50,thick, inner sep=0pt,minimum size=5mm},
tproj/.style={black,thick,<->},
tcont/.style={blue,thick,<->},
tcontb/.style={blue,thick,dashed,<->},
tbulk/.style={red,thin}
]
\node at (0,0) [tzop]  {$\rhob$} ;
\draw[tzstates] (-2.7,0) node[below=1.5mm] (b11)  {$\scriptstyle{\bra{v_a}}$} node[above=1.5mm] (k11) {$\scriptstyle{\ket{v_a}}$} -- ++(0.7,0.7) -- ++(-1.4,0) -- ++(1.4,-1.4) -- ++(-1.4,0) -- cycle;
\draw[tzstates] (2.7,0) node[below=1.5mm] (b12)  {$\scriptstyle{\bra{v_{\bar{a}}}}$} node[above=1.5mm] (k12) {$\scriptstyle{\ket{v_{\bar{a}}}}$} -- ++(0.7,0.7) -- ++(-1.4,0) -- ++(1.4,-1.4) -- ++(-1.4,0) -- cycle;
\draw[tproj] (-2.35,0.7) -- ++ (0,0.7) -- ++(4.7,0) -- ++(0,-0.7);
\draw[tproj] (-2.35,-0.7) -- ++ (0,-0.7) -- ++(4.7,0) -- ++(0,0.7);
\draw[tcont] (3.05,-0.7) -- ++ (0,-0.7) -- ++(0.7,0) -- ++(0,2.8) -- ++(-0.7,0) -- ++(0,-0.7);
\draw[tbulk]  (-2.7,0.7) -- ++ (0,0.4) -- ++(2.4,0) -- ++(0,-0.6);
\draw[tbulk]  (-2.7,-0.7) -- ++ (0,-0.4) -- ++(2.4,0) -- ++(0,0.6);
\draw[tbulk]  (2.7,0.7) -- ++ (0,0.4) -- ++(-2.4,0) -- ++(0,-0.6);
\draw[tbulk]  (2.7,-0.7) -- ++ (0,-0.4) -- ++(-2.4,0) -- ++(0,0.6);

\node at (9,0) [tzop]  {$\Ob$} ;
\draw[tzstates] (6.3,0) node[below=1.5mm] (b21)  {$\scriptstyle{\bra{v_a}}$} node[above=1.5mm] (k21) {$\scriptstyle{\ket{v_a}}$} -- ++(0.7,0.7) -- ++(-1.4,0) -- ++(1.4,-1.4) -- ++(-1.4,0) -- cycle;
\draw[tzstates] (11.7,0) node[below=1.5mm] (b22)  {$\scriptstyle{\bra{v_{\bar{a}}}}$} node[above=1.5mm] (k22) {$\scriptstyle{\ket{v_{\bar{a}}}}$} -- ++(0.7,0.7) -- ++(-1.4,0) -- ++(1.4,-1.4) -- ++(-1.4,0) -- cycle;
\draw[tproj] (6.65,0.7) -- ++ (0,0.7) -- ++(4.7,0) -- ++(0,-0.7);
\draw[tproj] (6.65,-0.7) -- ++ (0,-0.7) -- ++(4.7,0) -- ++(0,0.7);
\draw[tcont] (12.05,-0.7) -- ++ (0,-0.7) -- ++(0.7,0) -- ++(0,2.8) -- ++(-0.7,0) -- ++(0,-0.7);
\draw[tbulk]  (6.3,0.7) -- ++ (0,0.4) -- ++(2.4,0) -- ++(0,-0.6);
\draw[tbulk]  (6.3,-0.7) -- ++ (0,-0.4) -- ++(2.4,0) -- ++(0,0.6);
\draw[tbulk]  (11.7,0.7) -- ++ (0,0.4) -- ++(-2.4,0) -- ++(0,-0.6);
\draw[tbulk]  (11.7,-0.7) -- ++ (0,-0.4) -- ++(-2.4,0) -- ++(0,0.6);

\draw[tcontb] (-3.05,-0.7) -- ++ (0,-1) -- ++(7.5,0)  -- ++(0,3.0) -- ++(1.5,0) -- ++(0,-0.6);
\draw[tcontb] (-3.05,0.7) -- ++ (0,1) -- ++(8,0)  -- ++(0,-3.0) -- ++(1,0) -- ++(0,0.6);
\end{tikzpicture}
\caption{The representation of the computation of the r.h.s.\ of \eqref{eq:bdy1pt} which captures the non-trivial part of  $\Tr(\rho_A\, \mathcal{O}_A)$  in the random tensor network. We have two copies  of the network with swap boundary condition on $A$ (dashed blue) and identity boundary condition on $\bar{A}$ (solid blue). Note that the computation is similar to that involved in evaluating the second R\'enyi  entropy with one copy of the density matrix replaced by the operator.}
\label{fig:rtpetz}
\end{figure*}
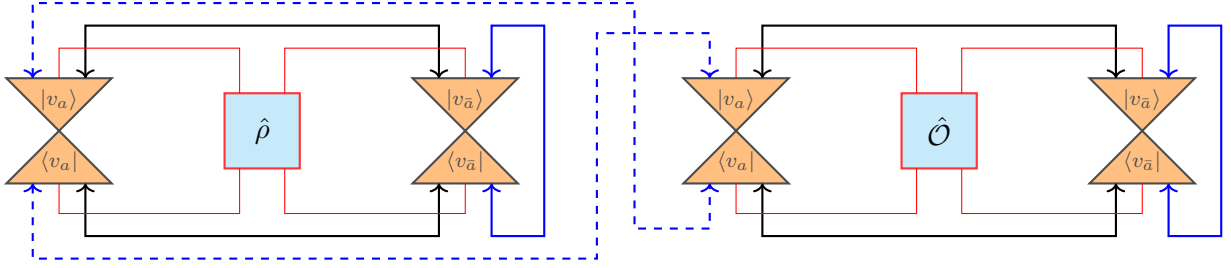

\begin{figure*}[ht]

\begin{subfigure}{\textwidth}
\centering
\begin{tikzpicture}
[scale=1.0,
tzop/.style={rectangle,draw=red!80,fill=cyan!20,thick, inner sep=0pt,minimum size=10mm},
tzstates/.style={black!70,fill=orange!50,thick, inner sep=0pt,minimum size=5mm},
tproj/.style={black,thick,<->},
tcont/.style={blue,thick,<->},
tcontb/.style={blue,thick,dashed,<->},
tbulk/.style={red,thin}
]

\node at (-1.5,0) {$\big\uparrow$};\node at (-1.5,-0.7) {$A$};
\node at (-0.5,0) {$\big\uparrow$};\node at (-0.5,-0.7) {$a$};
\node at (0.5,0) {$\big\downarrow$};\node at (0.5,-0.7) {$\bar{a}$};
\node at (1.5,0) {$\big\downarrow$};\node at (1.5,-0.7) {$\bar{A}$};

\draw[blue,thick,dashed] (-1.45,0) -- (-0.55,0);
\draw[black,thick] (-0.45,0) -- (0.45,0);
\draw[blue,thick] (0.55,0) -- (1.45,0);
\end{tikzpicture}
\caption{A simple network of 2 nodes  to illustrate the random average calculation. The spin configuration indicated  corresponds to the dominant contribution  in \cref{fig:paverage2}, with spin up representing  a swap operator and spin down represents an identity operator in the respective domains.}
\label{fig:spinmodel}
\end{subfigure}

\begin{subfigure}{\textwidth}
\centering
\begin{tikzpicture}
[scale=1.0,
tzop/.style={rectangle,draw=red!80,fill=cyan!20,thick, inner sep=0pt,minimum size=10mm},
tzstates/.style={black!70,fill=orange!50,thick, inner sep=0pt,minimum size=5mm},
tproj/.style={black,thick,<->},
tcont/.style={blue,thick,<->},
tcontb/.style={blue,thick,dashed,<->},
tbulk/.style={red,thin}
]
\node at (-6.25,0)  {$\overline{\Tr[\Mbb_A(\rhob)\otimes \Mbb_A(\Ob) \, X_A^{(2)}]} \;\; =  $};

\draw[tcontb] (-2.5,1) circle [radius =5mm]; \node at (-2,1.6) {2};
\draw[tproj] (-1,1) circle [radius =5mm] ;
\draw[tcont] (0.5,1) circle [radius =5mm] ; \node at (1,1.6) {2};
\draw [black] (-3,0) -- (1,0);
\node at (-1,-0.5) {$\chi^6$};
\node at (2,0) {$\times$};

\node at (4,1) [tzop]  {$\rhob$} ;
\node at (4,-0.7) [tzop]  {$\Ob$} ;
\draw[tbulk]  (4.3,1.5) .. controls (5,2.3) and (5,-0.3) .. (4.3,0.5);
\draw[tbulk]  (4.3,-0.2) .. controls (5,0.6) and (5,-2) .. (4.3,-1.2);
\draw[tbulk]  (3.7,0.5) -- (3.7,-0.2); 
\draw[tbulk]  (3.7,1.5) .. controls (2.8,2.6) and (2.8,-2.3) .. (3.7,-1.2);

\end{tikzpicture}
\caption{Illustration of the dominant contribution of the average over the random projectors, obtained by replacing $\ketbra{v_a}{v_a}^{\otimes 2}$ by swap operator and $\ketbra{v_{\bar{a}}}{v_{\bar{a}}}^{\otimes2}$ by identity operator in \cref{fig:rtpetz} for the simple network in \cref{fig:spinmodel}.  The result factorizes into a Bell pair term and a bulk term. For the Bell pair term each circle represents an index contraction loop that contributes a factor of $\chi$, leading to the domain wall term $e^{-|\gamma_A| \log \chi}$ (here $|\gamma_A| = 1$). The bulk term contains a non-trivial connection between $\rhob$ and $\Ob$ on $a$, leading to bulk one-point function $\Tr(\rhob_a \Ob_a)$. }
\label{fig:paverage2}
\end{subfigure}
\caption{The dominant contribution to random averaged expectation value illustrated in a 2-site network.}
\label{fig:paverage}
\end{figure*}

To verify that we have a good reconstruction, we will focus on evaluating the expectation values of the boundary operators. We have:
\begin{equation}\label{eq:bdy1pt}
\begin{split}
\Tr(\rho_A \, \mathcal{O}_A)  
&= 
	c_0\, \frac{\Tr[\Mbb_A(\rhob)\, \Mbb_A(\Ob) ]}{\Tr[\Mbb(\rhob)] }   \\
&= 
	c_0\, \frac{\Tr[\Mbb_A(\rhob)\otimes \Mbb_A(\Ob) \, X^{(2)}_A]}{\Tr[\Mbb(\rhob)] } \ 
\end{split}
\end{equation}	
where $X_A^{(2)}$ is the $\mathbb{Z}_2$ swap operator inserted at vertices in $A$.  The calculation pretty much parallels that of the second R\'enyi entropy, the main modification being that one of the copies of $\rhob$ is now replaced by the bulk operator $\Ob$, see \cref{fig:rtpetz}.

 It is not hard to see that  
\begin{equation}\label{eq:1ptnum}
\overline{\Tr[\Mbb_A(\rhob)\otimes \Mbb_A(\Ob) \, X_A^{(2)}]} = {\sf N}_2\,  c_0\, 
	e^{-\abs{\gamma_A} \,\log \chi } \, \Tr(\rhob_a \Ob_a)  + \cdots
\end{equation}	
where $\Ob_a = \Tr_{\bar{a}}(\Ob)$. In obtaining the answer we have assumed that the dominant spin configuration with the boundary conditions is $g_x = z_{(2)}$ for $x\in A$ and identity otherwise. A diagrammatic illustration of the averaged computation for a simple toy network is depicted in \cref{fig:paverage}.

The normalization factor from the denominator is computed similarly as before with the boundary vertices having the identity spin, leading to
\begin{equation}\label{eq:bdy1ptans}
\overline{\Tr(\rho_A \, \mathcal{O}_A)  }= \frac{{\sf N}_2} {{\sf N}_1}\,c_0\,  e^{-\abs{\gamma_A} \, \log \chi} \,  \Tr(\rhob_a \Ob_a)  + \text{subleading}\,.
\end{equation}	
The numerical pre-factor on the r.h.s.\ can be absorbed into the definition of the boundary operator $\mathcal{O}_A$ by  choosing $c_0$ appropriately. 
Thus, as expected the Petz-lite ansatz does a good job recovering the boundary operator $\mathcal{O}_A$ from the bulk data.

\subsection{General recovery: 1-point functions}
\label{sec:t1pt}

For the Petz map itself, we have to first make a choice of the fiducial reference state $\sigma$ and then take fractional powers. One can circumvent this by defining a replica version:
\begin{equation}\label{eq:Prep}
\mathcal{O}_A^{(n)} = \mathscr{E}(\sigma)^n \, \mathscr{E}\left(\sigma^{-n}\, \Ob \, \sigma^{-n}\right)\, \mathscr{E}(\sigma)^n \,,
\end{equation}	
and take the limit $n\to - \half$. While it is easy to see that this construction will work, specifically with the dominant spin configuration being a cyclic permutation for even $n$ (see for instance \cite{Penington:2019kki} who use a similar trick in the context of the SYK model), we find it useful to work with the twirled Petz map. As recovery channel, the twirled Petz map is given by the expression:
\begin{equation}
\mathscr{R}_\sigma(\rhob) = \int_{-\infty}^\infty\, \frac{dt}{1+\cosh t } \frac{\pi}{2} \, \sigma^\frac{1-it}{2}\, \mathscr{E}^\dagger\left(
[\mathscr{E}(\sigma)]^{-\frac{1-it}{2}}\, \rhob\, [\mathscr{E}(\sigma)]^{-\frac{1+it}{2}}
\right) \sigma^\frac{1+it}{2}
\end{equation}	
for an arbitrary reference state $\sigma$.  Picking the reference state to the maximally mixed state $\tau  =\frac{1}{d_\text{code}}\, \Idop_\text{code} $, one can write an expression for the operator version of the map which allows a more standard replica construction. One has \cite{Cotler:2017erl,Chen:2019gbt}
\begin{equation}\label{eq:twistedP}
\begin{split}
\mathcal{O}_A &= \frac{1}{d_\text{code}} \, \dv{}{t}\bigg|_{t=0}\, \log\left[\Mbb_A\left( \sigmab(t) \right)\right]\\
\sigmab(t) &\equiv \frac{1}{d_\text{code}}\, \Idop_\text{code} + t\, \Ob_a \otimes \Idop_{\bar{a}} \,,
\end{split}
\end{equation}	
where $d_\text{code} = d_a \times d_{\bar{a}}$.  In  this presentation of the twirled Petz  map, we can think of $t$ as quantifying a perturbation of the maximally mixed state -- it may  be viewed as a source for the operator deformation by $\Ob_a \otimes \Idop_{\bar{a}} $.

The representation of the twirled Petz map in \eqref{eq:twistedP} has the advantage that we can use the standard replica trick used to compute relative entropy. Using the standard identify
\begin{equation}\label{eq:reprel}
\Tr\left(\rho \log \sigma\right) = \lim_{n\to 1} \, \frac{1}{n-1}	 \log\Tr\left(\rho\,\sigma^{n-1}\right) ,
\end{equation}	
we can rewrite an insertion of $\mathcal{O}_A$ in terms of insertions of powers of $\Mbb(\sigmab(t))$ which will allow for direct replica manipulations, as we illustrate below.

To test the efficacy of the reconstruction we again focus on matching expectation values of operators in the bulk and boundary states respectively. We have 
\begin{equation}\label{eq:twistedPetzrep}
\begin{split}
\Tr\left(\rho_A\, \mathcal{O}_A\right) 
&=
	\frac{1}{d_\text{code}} \, \dv{}{t}\,  \lim_{n\to 1} \, \frac{1}{n-1} \, \log\mathfrak{f}_{(n)}(t) \bigg|_{t=0}\\\
\mathfrak{f}_{(n)}(t) 
&\equiv \Tr\left[ \Mbb_A(\rhob) \, \Mbb_A^{n-1}(\sigmab(t))\right] .
\end{split}
\end{equation}	
We have chosen to drop a normalization factor $\Tr\left[\Mbb(\rhob)\right]$ above as it independent of our deformation parameter $t$.
We can furthermore factorize the computation of $\mathfrak{f}_{(n)}(t)$ by realizing that we can replace $\Mbb_A(\sigmab)^{n-1}$ by 
$\Mbb_A(\sigmab)^{\otimes (n-1)} \, X^{(n)}_A$, as can be pictorially visualized as before.

The computation is again simplified if we first perform the averaging over the random projectors and proceeds in parallel to our earlier discussion for the computation of the R\'enyi entropies. We have 
\begin{equation}
\overline{\mathfrak{f}_{(n)}(t)} = {\sf N}_{n} \,e^{-(n-1)\, \abs{\gamma_A}\, \log\chi} \, \Tr\left(\rhob_a\, \sigmab^{n-1}_a\right) + \cdots
\end{equation}	
with 
\begin{equation}
\begin{split}
\sigmab_a &= \Tr_{\bar{a}} \sigmab(t) = \frac{1}{d_a} + t\, d_{\bar{a}}\, \Ob_a \,,
\end{split}
\end{equation}	
which implies
\begin{equation}
	\Tr(\rhob_a\, \sigmab_a^{n-1} )  
	= \frac{\Idop_a}{d_a^{n-1}} + t\, \frac{(n-1)\, d_{\bar{a}}}{d_a^{n-2}}\, \Tr(\rhob_a\, \Ob_a) + \order{t^2}\,.
\end{equation}	
Differentiating with respect to the deformation parameter $t$ and then taking the limit $n\to1$ we end up the desired answer 
\begin{equation}
\overline{\Tr(\rho_A\, \mathcal{O}_A)} = \Tr(\rhob\, \Ob) +\text{subleading} \,.
\end{equation}	
%

\subsection{General recovery: higher point functions}
\label{sec:thp}

Having understood the Petz reconstruction of bulk operators, and the recovery of the 1-point functions, we now turn to the higher point functions. The general statement of entanglement wedge reconstruction would say that we should be able to recover an arbitrary correlation function of bulk operators in the entanglement wedge in terms of corresponding boundary avatars. It was already argued in \cite{Cotler:2017erl} that the Petz reconstruction would achieve this outcome. We will now verify the same explicitly in the RTNs.

The recovery  of an $k$-point function using the Petz-lite reconstruction map is a straightforward generalization of the computation in \cref{sec:t1pt}. One can immediately see that it parallels the computation of the $k+1$ R\'enyi entropy where $k$ copies of $\Mbb(\rhob)$ are replaced by  $\prod_{j=1}^k \, \Mbb(\Ob_j)$. We will therefore focus on recovering the higher-point functions using the twirled Petz map.  For the sake of illustration consider first the computation of the two-point function of operators inserted in  the homology surface $a$. We want to show that 
\begin{equation}\label{eq:2pttarget}
\expval{\mathcal{O}_{A1} \, \mathcal{O}_{A2}}_{\rho} = \expval{\Ob_{a1}\, \Ob_{a2}}_{\rhob} + \text{subleading} \,.
\end{equation}	

We will use the replica trick to compute the l.h.s.\ of \eqref{eq:2pttarget}, for which we need a suitable generalization of \eqref{eq:reprel} 
when multiple operators are in play. Consider therefore the following:
\begin{equation}\label{eq:reprel2}
\begin{split}
\expval{\log \sigma_1\, \log\sigma_2}^c_{\rho} 
&\equiv
	\Tr(\rho\, \log \sigma_1 \, \log \sigma_2)- \Tr(\rho \log \sigma_1)\, \Tr(\rho \,\log \sigma_2)\\
&=
	\lim_{n_1 \to 1} \lim_{n_2 \to 1} \, \frac{1}{n_1-1} \, \frac{1}{n_2-1} \, \log \Tr(\rho\, \sigma_1^{n_1-1}\, \sigma_2^{n_2-1}) .
\end{split}
\end{equation}	
One can justify this in a manner similar to \eqref{eq:reprel}. To be clear, one can use a unitary diagonalization ansatz $\sigma_i = U_i\,\Sigma_i\, U^\dagger_i $ to  facilitate computation of the powers, and then differentiates with respect to $n_1$ and $n_2$ to arrive at the limit. The reason for ending  up  with the connected part of the correlator is the usual fact that the logarithm isolates the connected pieces (sequential derivatives give the lower-point functions).

With the multi-replica trick \eqref{eq:rtnreplica2} at hand one can proceed with the computation of the two-point function of the reconstructed operators. We have:
\begin{equation}
\expval{\mathcal{O}_{A1} \, \mathcal{O}_{A2}}_{\rho}^c 
=
	\left( \frac{1}{d_\text{code}}\right)^2 \, \lim_{n_1,n_2\to 1}  \pdv{t_1} \, \pdv{t_2}  \prod_{i=1}^2\, \frac{1}{n_i-1}\,\log \mathfrak{f}_{(n_1,n_2)}(t_1,t_2) \bigg|_{t_1=t_2=0}\,,
\end{equation}	
where 
\begin{equation}
\mathfrak{f}_{(n_1,n_2)}(t_1,t_2) = \Tr[\Mbb_A(\rhob)\, \Mbb_A^{n_1-1}(\sigmab(t_1)) \, \Mbb_A^{n_2-1} (\sigmab(t_2))] .
\end{equation}	
Once again we can unfold the computation to work in a tensor product replica space with suitable cyclic permutations. Taking the average over the random projectors allows us to evaluate the result. One finds: 
\begin{equation}
\overline{\mathfrak{f}_{(n_1,n_2)}(t_1,t_2) }
	= {\sf N}_M\, e^{-(M-1)\, \abs{\gamma_A} \, \log\chi} \, \Tr[\rhob_a \prod_{i=1}^2 \left(\frac{1}{d_a^{n_i-1}} 
	+ t_i\, \frac{(n_i-1)\,d_{\bar{a}}}{d_a^{n_i-2}}  \, \Ob_{ai}\right)] + \mathcal{O}\left( t^2_i \right) \,,
\end{equation}	
with $M = 1+\sum_i\, (n_i-1)$.  Finally, we can evaluate the derivatives with respect to $t_i$ and thence the limit $n_i \to 1$ recovering for the average:
\begin{equation}\label{eq:2ptrec}
\overline{\expval{\mathcal{O}_{A1} \, \mathcal{O}_{A2}}_{\rho}^c }= \expval{\Ob_{a1}\, \Ob_{a2}}_{\rhob}^c + \text{subleading}\,.
\end{equation}	
Effectively, the random projection average allows us to interpret  $\overline{\mathfrak{f}_{(n_1,n_2)}(t_1,t_2) }$ as the generating functional of the correlation functions with $t_1$ and $t_2$ as sources. So indeed $\log \overline{\mathfrak{f}_{(n_1,n_2)}(t_1,t_2) }$ does lead to connected 2-point  correlators. Since we independently have a computation of the expectation values of one-point functions, we can immediately recover the full two-point function. 

The matching of higher point functions works similarly, once we realize that we can write:
\begin{equation}\label{eq:relrepn}
\expval{\prod_{j=1}^k\,\log \sigma_j }_{\rho}^c =\left( \prod_{i=1}^k \lim_{n_i \to 1} \frac{1}{n_i-1}\right)  \log \Tr(\rho \prod_{j=1}^k \,\sigma_j^{n_j-1}) .
\end{equation}	
Thence a similar reasoning as above leads us to conclude that 
\begin{equation}\label{eq:nptrec}
\overline{\expval{\prod_{j=1}^k\,\mathcal{O}_{Aj}} _{\rho}^c }= \expval{\prod_{j=1}^k\, \Ob_{aj}}_{\rhob}^c + \text{subleading}\,.
\end{equation}	
Recovering the connected parts suffices as the disconnected parts can be reconstructed iteratively  from the lower-point functions.

\section{Comments on holographic channels}
\label{sec:comments}

We have focused on illustrating the efficacy of the Petz reconstruction of bulk operators in RTNs by concentrating on the matching of correlation functions. We see that  the general correlation functions of bulk operators in the homology region $R_A$ in the reduced state $\rhob_a$  agree with those computed using the corresponding reconstructed boundary operators in accord with \cite{Cotler:2017erl}.

The underlying reason for the matching is simply the fact that we are able to perform state decoding. That is, given $\rho_A = \mathscr{E}(\rhob)$, the recovery map viewed as  a decoding channel satisfies Eq.~\eqref{eq:petz}, viz.,  $\mathscr{R}\circ\mathscr{E}(\rhob) \approx \rhob_a$. Indeed this suffices, as
\begin{equation}
\Tr_A(\mathcal{O}_A \, \rho_A) = \Tr_A(\mathscr{R}^\dagger(\Ob_a) \, \rho_A) = \Tr_a(\Ob_a\, \mathscr{R}(\rho_A)) 
	= \Tr_a(\Ob_a\, \rhob_a)\,.
\end{equation}	
The state decoding itself works as follows. Given a bulk state $\rhob$ and its boundary encoding $\mathscr{E}(\rhob)$ the random average decoded state is 
\begin{equation}
\overline{\mathscr{R}\circ\mathscr{E}(\rhob)} = \rhob_a \otimes \mathbf{1}_{\bar{a}} + \cdots\,.
\end{equation}	
One can read this off directly from \cref{fig:rtpetz} -- the construction simply involves erasing the operator $\Ob$ and leaving the bulk legs on the second copy dangling to create a state in $a$ as we expect from the general results proved in \cite{Cotler:2017erl,Chen:2019gbt}.

However, as noted above the bulk-boundary map in RTNs is not trace-preserving, and thus not a quantum channel per se, but rather a quantum operation.  One can nevertheless proceed with an analog of a Petz reconstruction map which (as demonstrated above) serves to reconstruct the bulk operators. 

The crucial fact that $\Ob_a$ lies in the homology surface $R_A =a$ was left implicit in our analysis, but follows from the nature of the dominant (saddle-point) spin configuration minimizing the energy in the auxiliary spin model. For instance, if we focus on a simplified toy problem where we have a single bulk degree of freedom in the middle, then we can illustrate the operation $\overline{\mathscr{R}\circ\mathscr{E}}$ as a quantum depolarizing operation. One can write the averaged decoding map as
\begin{equation}
\overline{\mathscr{R}\circ\mathscr{E}(\rhob)}  \propto e^{-\abs{\gamma_A^1} \, \log\chi} \, \rhob + e^{-\abs{\gamma_A^2} \, \log \chi} \, \hat{\sigma} \,.
\end{equation}	
We have now written explicitly the result with two potential energy minimizing domain wall configurations. For $\chi \gg 1$, we have a sharp phase transition between two situations; when $\gamma_A^1$ is the minimum cut, then the recovery succeeds with certainty. On the other hand when $\gamma_A^2$ is the minimum cut the recovery fails completely as we end up with the maximally mixed state.

One key aspect of the RTNs that is worth emphasizing has to do with the use of random projectors to define the bulk to boundary map. For any graph network, the boundary encoding $\Mbb_A(\rhob)$ can be well approximated by averaged boundary density operator $\overline{\Mbb_A(\rhob)}$, as one would expect. However, computation of moments of the density operators, or the evaluation of correlation functions, as discussed in the preceding sections, involves non-trivial interlinking across the replica copies. Mathematically, this is clear from the fact that such computations in a single theory can be unfolded into an evaluation in the $n$-fold tensor product system with suitable insertion of cyclic permutation elements. When we carry out the average over the random projectors, we can induce connections between these replica copies as the group averaging result \eqref{eq:projavg} provides further permutation insertions which can combine with the cyclic replica permutation to generate  new cycles/links. 

This is entirely analogous to the manner in which gravitational dynamics engenders connections between different replica copies through Euclidean replica wormholes \cite{Penington:2019kki,Almheiri:2019qdq} as is to a large extent already clear from the results of \cite{Penington:2019kki} on Petz reconstruction.   While one might have viewed the random tensors as a means to extract the behaviour of the typical state of the graph, the general lesson is that the fluctuations, moments, and correlations in such random projected states,  carry  
information beyond the average, thanks to the aforementioned propensity for forming new inter-connections.

In this vein it is interesting to contemplate the distinction between coarse-graining (via partial tracing) versus random projections onto entangled states for general open quantum systems. In a bipartite system-environment setting, one traditionally considers simply tracing out environmental degrees of freedom, inducing on the system degrees of freedom a non-unitary dynamics. However, the results alluded to above suggest  that one may be able to glean further insight from the study of random projection models of system-environment couplings. More specifically, in these settings, replicaesque analysis ought to reveal that the system has more detailed information about the purifying environment than would have been anticipated naively. These issues deserve further attention.

\acknowledgments 
It is a pleasure to thank  Veronika Hubeny and Edward Witten for useful discussions and  valuable feedback. 
This work was  supported by U.S.\ Department of Energy grant {DE-SC0019480} under the HEP-QIS QuantISED program and by funds from the University of California.

\bibliographystyle{JHEP}

\providecommand{\href}[2]{#2}\begingroup\raggedright\endgroup

\end{document}